\documentclass[english,aps,prx,showpacs,citeautoscript,twocolumn]{revtex4}
\usepackage{amsmath}
\usepackage{amssymb}
\usepackage{graphicx}
\usepackage{array}
\usepackage{float}
\usepackage{mathtools}
\usepackage{longtable}
\LTcapwidth=\textwidth

\date{\today}

\begin{document}

\title{Machine learning the band gap properties of kesterite I$_2$-II-IV-V$_4$ quaternary compounds for photovoltaics applications}

\author{L.\ Weston$^{1,2}$ and C. Stampfl$^1$}
\affiliation{$^1$School of Physics, The University of Sydney, Sydney, New South Wales 2006, Australia}
\affiliation{$^2$Materials Department, University of California, Santa Barbara, California 93106-5050, USA}

\begin{abstract}
Kesterite I$_2$-II-IV-V$_4$ semiconductors are promising solar absorbers for photovoltaics applications. The band gap and it's character, either direct or indirect, are fundamental properties determining photovoltaic-device efficiency. We use a combination of accurate first-principles calculations and machine learning to predict the properties of the band gap for a large number of kesterite I$_2$-II-IV-V$_4$ semiconductors. In determining the magnitude of the fundamental gap, we compare results for a number of machine-learning models, and achieve a root mean squared error as low as 283 meV; the best results are achieved using support-vector regression with a radial-bias kernel. This error is well within the uncertainty of even the most advanced first-principles methods for calculating semiconductor band gaps. Predicting the direct--indirect property of the band gap is more challenging. After significant feature engineering, we are able to train a classifier that predicts the nature of the band gap with an accuracy of 89 \% using logistic regression. Using these trained models, the band gap properties of 1568 kesterite I$_2$-II-IV-V$_4$ compounds are predicted. We find 717 compounds with band gaps in the range 0.5 -- 2.5 eV that can potentially act as solar absorbers, and 242 materials with a band gap in the ``\emph{optimum range}" of 1.2 -- 1.8 eV. The stability of these 242 compounds is assessed by calculating the Energy Above Hull using the Materials Project database, and the band gaps are verified using hybrid functional calculations; in the end, we identify 25 compounds that are expected to be synthesizable, and have a band gap in the range 1.2 -- 1.8 eV -- most of which are previously unexplored. These results will be useful in the materials engineering of efficient photovoltaic devices.
\end{abstract}

\pacs{71.20.-b, 61.66.Dk, 42.79.Ek}

%71.20.-b 	Electron density of states and band structure of crystalline solids
%61.66.Dk 	Alloys
%42.79.Ek 	Solar absorers

\maketitle

\def\figurename{FIG.}\def\tablename{TABLE}

\section{Introduction}
Quaternary I$_2$-II-IV-VI$_4$ semiconductors offer a unique opportunity in materials engineering  due to the vast design space \cite{CaiCM2015, ChenPRB2009}. The different cation--anion combinations exhibit band gaps spanning the visible spectrum.
This tunability in the band gap upon cation and anion mutation has lead to intense interest in these materials for applications as solar absorbers for photovoltaic devices \cite{ItoJJAP1988, KatagiriAPE2008, SiebentrittPP2012}.

For photovoltaics applications, the band gap is a fundamental property determining efficiency, with band gaps around 1.5 eV being the most efficient solar absorbers \cite{RuhleSE2016}. Moreover, the direct--indirect character of the band gap is of a fundamental importance: while direct gap materials are typically stronger absorbers than indirect materials, they may also have shorter photocarrier lifetimes and suffer from carrier recombination \cite{NelsonBOOK2003}.
Given the complex design space of I$_2$-II-IV-VI$_4$ compounds, it becomes difficult to characterize all of the possible cation--anion combinations, both theoretically and experimentally. There exists multiple possibilities for both the cation ordering, including kesterite and stannite, as well as the crystal symmetry, since the geometry may be derived from either the zinc blende or wurtzite phase \cite{ChenPRB2010}; consequently, there are thousands if not tens of thousands of possible I$_2$-II-IV-VI$_4$ materials. 

From a theoretical perspective, the calculation of semiconductor band gaps within traditional density functional theory (DFT) suffers from the well-known underestimation error \cite{MoriPRL2008}. This can be overcome with more accurate theoretical approaches such as screened hybrid functionals \cite{HeydJCP2005} or many-body perturbation theory ($GW$) \cite{ShishkinPRB2007}. However, these are far more computationally expensive, and therefore are difficult to implement on a large set of materials.  Indeed, the available large databases of semicondcutor band gaps mostly rely on traditional DFT calculations within the generalized-gradient approximation (GGA) \cite{CurtaroloCMS2012, JainAPL2013, SaalJOM2013}.

One possible approach to overcome this challenge, is to use machine learning to generate or improve predictions \cite{PilaniaSR2013, RamakrishnanJCTC2015}. By performing accurate high-level first-principles calculations  on a subset of I$_2$-II-IV-VI$_4$ compounds, the results can be used to train a machine-learning model to predict the properties of the remaining materials in the design space. Lee \emph{et al}.\ used machine learning to predict the band gaps of 156 $A$X binary compounds using element-specific descriptors including the band gap from low-level DFT calculations, achieving a root mean squared error (RMSE) of 180 meV with support-vector reggression \cite{LeePRB2016}. Pilania \emph{et al}.\ used kernel-ridge regression to predict the band gaps of 1306 double perovskites and achieved a RMSE of 80 meV using a 16-dimensional set of element-specific descriptors \cite{PilaniaSR2016}. Ward \emph{et al}.\ proposed a large set of 140 universal descriptors to predict band gaps from a very large data set, and identified new possible solar absorbers \cite{WardNCM2016}; it was also found that model accuracy was improved by partitioning the data set into groups of similar materials, suggesting that machine-learning predictions would work best on isostructural and isoelectronic materials. To the best of our knowledge, classification of band gaps as either direct or indirect has not been attempted from a machine-learning perspective.

In the present paper, we study I$_2$-II-IV-VI$_4$ semiconductors in the zinc-blende-based kesterite structure. This provides the opportunity to study a large number of materials systems that are both isostructural and isoelectronic. We consider 1568 possible cation--anion combinations, and perform accurate hybrid functional calculations for the band gaps on a randomly-selected subset of 200 materials; these results are then used to train various machine-learning models. Using support-vector regression with a radial bias kernel, we are able to predict the magnitude of the fundamental gap with a RMSE of 283 meV, using only 3 simple, element-specific descriptors per element in the compound. We find that classification of these materials as direct or indirect semiconductors is more challenging. After substantial feature engineering, we train a classifier with an accuracy of 89\% using logistic regression.
The trained models are used to predict the band gap properties for all 1568 compounds, and to identify potential solar absorbers; these results will be useful in the design and engineering of kesterite I$_2$-II-IV-VI$_4$ semiconductors for photovoltaics applications.

\section{Methodology}
\subsection{Materials systems}
\begin{figure}
\centering
\includegraphics[width=6cm]{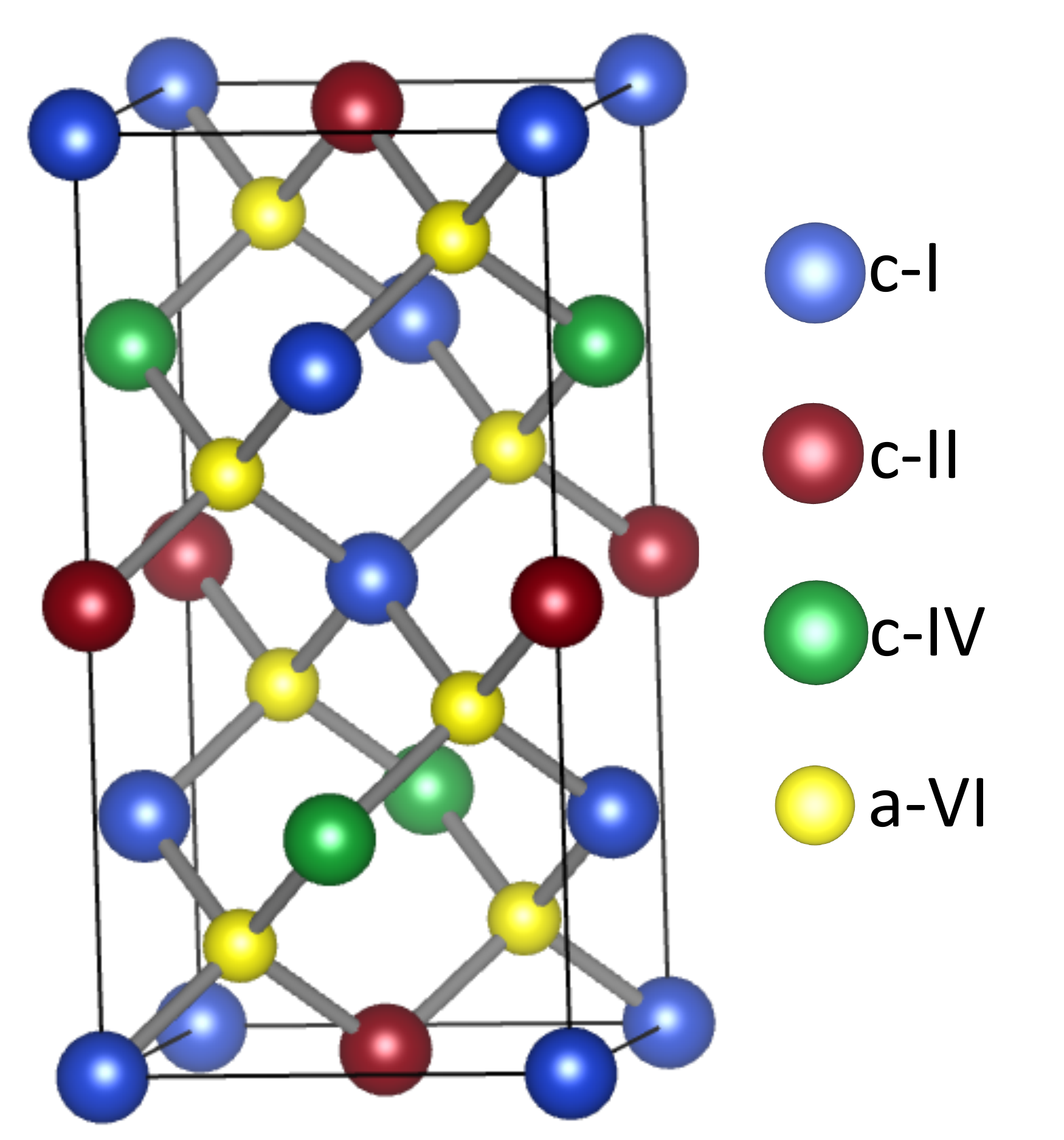}
\caption{The zinc-blende-based kesterite structure for a  I$_2$-II-IV-VI$_4$ compound. The cations with valence I (c-I), II (c-II), and IV (c-IV) are indicated by blue, red, and green spheres, respectively. The anion has a valence VI (a-VI) and is indicated by a yellow sphere.}
\label{fig1}
\end{figure}

The kesterite structure is derived via cation mutation of the II-VI binary zinc blende phase, and is shown in Fig.~\ref{fig1}. For the I$_2$-II-IV-VI$_4$ compounds, we consider: I = Li, Na, K, Rb, Cs, Cu, Ag; II = Be, Mg, Ca, Sr, Ba, Zn, Cd, Hg; IV = C, Si, Ge, Sn, Ti, Zr, Hf; and, VI =  O, S, Se, Te. This provides a total of  1568 compounds. While a number of these compounds will not be thermodynamically stable, they are still useful in training the machine-learning models.  We randomly select a subset of 200 materials, and calculate their band gap properties.

\subsection{First-principles calculations}
Our calculations are performed in the Vienna \emph{Ab initio} Simulation Package (VASP) \cite{KressePRB1996}, using density functional theory (DFT) within the generalized Kohn-Sham scheme \cite{KohnPR1965}. The valence electrons are separated from the core by use of projector augmented wave (PAW) potentials \cite{BlochlPRB1994}.

The lattice parameters and the internal ionic coordinates are determined by a full relaxation of the cell using the PBEsol functional \cite{PerdewPRL2008}; PBEsol has been shown to give highly accurate geometries for zinc blende semiconductors \cite{LuceroJP2012}. Once the geometry has been determined, we perform a fixed-point calculation of the band gap using the screened hybrid functional of Heyd, Scuseria and Ernzerhof (HSE) \cite{HeydJCP2003, HeydJCP2006}. In this approach, the short-range exchange potential is calculated by mixing a fraction of non-local Hartree-Fock exchange with the GGA of Perdew, Burke and Ernzerhof (PBE) \cite{PerdewPRL1996}. The long-range exchange potential and the correlation potential are calculated with PBE. The screening parameter is set to 0.2 \AA$^{-1}$ and the mixing parameter to $\alpha = 0.25$. The HSE functional provides highly accurate semiconductor band gaps when compared to traditional DFT \cite{HeydJCP2005}. For the stability analysis in Section.~IIIC, we use the PBE functional in the formation enthalpy calculations to be more consistent with the Materials Project database \cite{JainAPL2013}.

The kesterite phase has an 8-atom body-centered tetragonal (BCT) primitive cell. For the geometry relaxation within PBEsol, an $8\times8\times8$ Monkhorst-Pack $k$-point grid is used for integrations over the Brillouin zone \cite{MonkhorstPRB1976}. For determination of the band gap within HSE, we perform a full calculation along the high-symmetry path in the BCT Brillouin zone \cite{SetyawanCMS2010}. We use a plane wave cutoff of 400 eV for the sulfides, selenides and tellurides, and 500 eV for the oxides. For the selenides and tellurides, the spin-orbit splitting ($\Delta_{SO}$) at the valence band maximum is neglected; however, the splitting only affects the band gap by $\Delta_{SO}/3$, and therefore we expect an error less than 100 meV in most cases \cite{ChenAPL2009}.

\subsection{Machine-learning models}
\subsubsection{Regression}

The magnitude of the band gap can be predicted using a number of regression models. Regression aims to determine a relationship between the features of each compound, called descriptors (discussed below), and the band gap of the material. We present the key feature of each model below.

\noindent \emph{Linear and support-vector regression}.
Consider a linear function $y = \left< \omega, x\right> + b$, where $\omega$ and $x$ are vectors and $\left<.,.\right>$ denotes a dot product; for a set of features $x_i$ and outcomes $y_i$, an ordinary least squares regression will attempt to fit $\omega$ and $b$ to minimize the sum of squares $\sum_i \left[y_i - \left(\left<\omega, x_i\right> + b \right)\right]^2$. Support-vector regression introduces the concept of a margin $\varepsilon$, and attempts to fit a curve such that all of the points lie within the margin. Support-vector regression also favours curve ``\emph{flatness}", by reducing the sensitivity to outliers. The problem of support-vector regression is typically written in the following way \cite{SmolaSC2004},
\begin{flalign}
\mathrm{minimize}&: \ ||\omega^2|| \\\nonumber
\mathrm{subject \ to}&: \ | y_i - \left(\left<\omega, x_i\right> + b\right) | \leq \varepsilon.
\end{flalign}
Here, minimizing $||\omega^2||$ maximizes the \emph{flatness} in the curve.  In many cases, it is not possible to fit a curve such that $| y_i - (\left<\omega, x_i\right> + b) | \leq \varepsilon$, and therefore additional parameters are introduced to construct a so-called ``\emph{soft margin}''. There is a trade-off between the \emph{softness} in the margin and the \emph{flatness} in the curve that is determined by a constant known as the $C$ parameter, and this must be tuned to optimize predictions.

In addition to linear support-vector machines, a non-linear transformation may be applied on the feature space by the so-called ``\emph{kernel trick}". We implement support-vector regression with a radial bias function. For two data points $x$ and $x'$, this function is defined as follows,
\begin{eqnarray}
R(x, x') = \exp \left(-\gamma ||x - x'||\right) .
\end{eqnarray}
The parameter $\gamma$ determines how quickly $R$ decays with the distance between $x$ and $x'$ in the feature space; $\gamma$ can be tuned to optimize predictions.

\noindent \emph{Tree-based methods}.
The most simple tree-based regression is a decision tree regressor \cite{BreimanBOOK1984}. For a set of inputs $x_i$ and outcomes $y_i$, the decision tree will split regions in the feature space into two groups, $R_1$ and $R_2$, having mean outcomes $\widehat{y_1}$ and $\widehat{y_2}$, in such a way that minimizes the residual sum squared ($RSS$),
\begin{eqnarray}
RSS = \sum_{i \in R_1} (y_i - \widehat{y_1})^2 + \sum_{i \in R_2} (y_i - \widehat{y_2})^2 .
\end{eqnarray}
After the initial split, the tree will continue to make further optimum splits in the feature space until some convergence criteria is met, and the remaining unsplit groups $R_t$ are called terminal nodes, or leaves. The total function to minimize for the decision tree regressor ($DTR$) is the following, 
\begin{eqnarray}
DTR = \sum_{t = 1}^{T} \sum_{i \in R_t} (y_i - \widehat{y_t})^2 + \gamma T, \label{eq:residuals}
\end{eqnarray}
where $T$ is the number of terminal nodes, and $\widehat{y_t}$ is the mean of all $y_i$ in $R_t$. The second term $\gamma T$ is a penalty that prevents over fitting for more complex trees.

Decision trees are computationally efficient and easy to interpret, however often suffer from over fitting and inaccurate predictions. Ensemble-tree based methods are typically used to overcome the shortcomings of simple decision trees. One such method is the random forest regressor \cite{LiawRN2002}. Random forest regression works using the technique of bootstrap aggregating, in which a random subset of $x_i$ and $y_i$ are chosen to train a decision tree; this process is repeated to fit many trees, and then predictions can be made by averaging the results from the ensemble of regression trees.  The number of randomly selected decision trees to be fitted is a parameter that is typically tuned to optimize predictions. Another ensemble method is the gradient-boosted regression tree \cite{ElithJAE2008}. Boosting is a technique in which many individual decision trees are trained sequentially; each tree is trained from the residuals of the previous tree, as defined by Eq.~\ref{eq:residuals}. In this way, the new tree that is added to the ensemble is the one that best minimizes the residuals.
Ensemble methods such as random forest and boosting typically correct for over fitting, and reduce the sensitivity to noise in the training set.

\subsubsection{Classification}
We train a binary direct--indirect band gap classifier using logistic regression. In this approach, the model can predict the probability of a binary outcome, and makes predictions on the outcome by determining which is more likely. The logistic function $L(x)$ is an $S$-shaped curved that varies smoothly between 0 and 1; $L(x)$ takes the vector of features $x$, and if $L(x) < 0.5$, the classifier will predict a binary outcome of 0; when $L(x) > 0.5$, the classifier predicts a binary outcome of 1.
Similar to linear regression, which attempts to fit the optimum coefficients for the linear equation $\left< \omega, x\right> + b$, logistic regression is fit by optimizing the coefficients in $L(x)$, 
\begin{eqnarray}
L(x) = \frac{1}{1+\exp \left(\left[-(\left< \omega, x\right> + b\right)\right]} ,\label{eq:log}
\end{eqnarray}
by minimizing the number of incorrect classifications on the training data. 
In the present study, the binary outcomes are direct--indirect rather than 0--1.

\subsubsection{Feature space}
\label{sec:features}
A number of different features have been proposed as predictors for materials properties \cite{FaberPRL2016, LeePRB2016, PilaniaSR2016, WardNCM2016}. In the present work, we first use a simple set of element-specific features. For each of the elements in the I$_2$-II-IV-VI$_4$ compound, the electronegativity, ionic radius, and row in the periodic table are used; this gives 12 features total per compound. This 12-dimensional feature space works extremely well for predicting the magnitude of the band gap using regression techniques. However, this set of features performed poorly when implemented in the direct-indirect band gap classifier, and we had to perform substantial feature engineering, as will be discussed in in Sec.~\ref{sec:class}.

\section{Results \& Discussion}
Of the 200 compounds studied, 16 either did not have a band gap, or did not converge at some stage of the calculation; these were excluded from the fitting. 
The band gaps of the remaining 184 I$_2$-II-IV-VI$_4$ compounds are used to train the machine-learning models. We first discuss determination of the magnitude of the fundamental gap using regression models. Next, we will discuss training of a classifier to determine the direct--indirect character of the gap.

\subsection{Band gap regressor}
A number of regression models are used to fit the magnitude of the band gap. Where appropriate we performed feature normalization, and performed a search over any tunable parameters to optimize the regressor. The accuracy of the model is determined using 10-fold cross validation.  The accuracy of the model is assessed by analyzing the root mean squared error (RMSE), and the $R^2$ coefficient of determination. The results for each regression model are presented in Table~\ref{tab:error}.

\begin{table}
\setlength{\tabcolsep}{14pt}
\setlength{\extrarowheight}{4.5pt}
\begin{tabular}{lccccc} \hline\hline
 Model   &  $R^2$ & RMSE (eV)   \\ \hline
Linear Regression & 0.796 &  0.590  \\
SVR-L   &   0.789     &   0.592  \\
SVR-RBF   &   0.957   &  0.283  \\
Decision Tree  &  0.823 & 0.492 \\ 
Random Forest  &  0.874  & 0.435 \\ 
Boosted Reg. Tree  &  0.934  & 0.358 \\ \hline\hline
\end{tabular}
\caption{Root mean squared error (RMSE) and $R^2$ value for machine-learning models based on 10-fold cross validation. Results are shown for linear regression, support-vector regression with a linear (SVR-L) and radial bias function (SVR-R) kernel, decision tree, random forest and boosted  regression tree (Boosted Reg. Tree).}
\label{tab:error}
\end{table}	

\begin{figure}
\centering
\includegraphics[width=8.6cm]{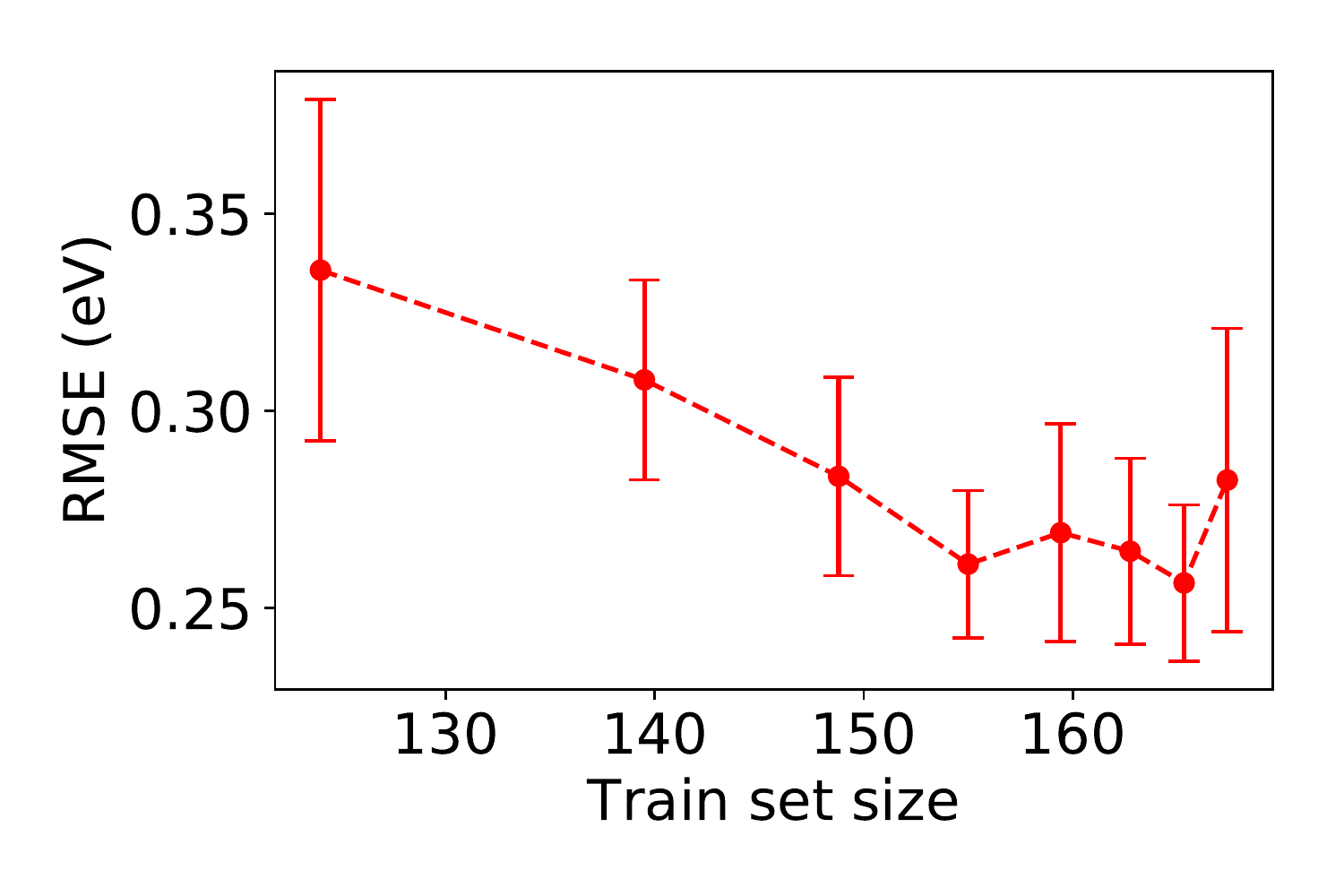}
\caption{For a nonlinear support vector machine, the root mean squared error (RMSE) is plotted for the band gap predictions as a function of training set size. The error bars represent the standard deviations in the RMSE from n-fold cross validation.}
\label{fig_sample_size}
\end{figure}

\begin{figure}
\centering
\includegraphics[width=8.6cm]{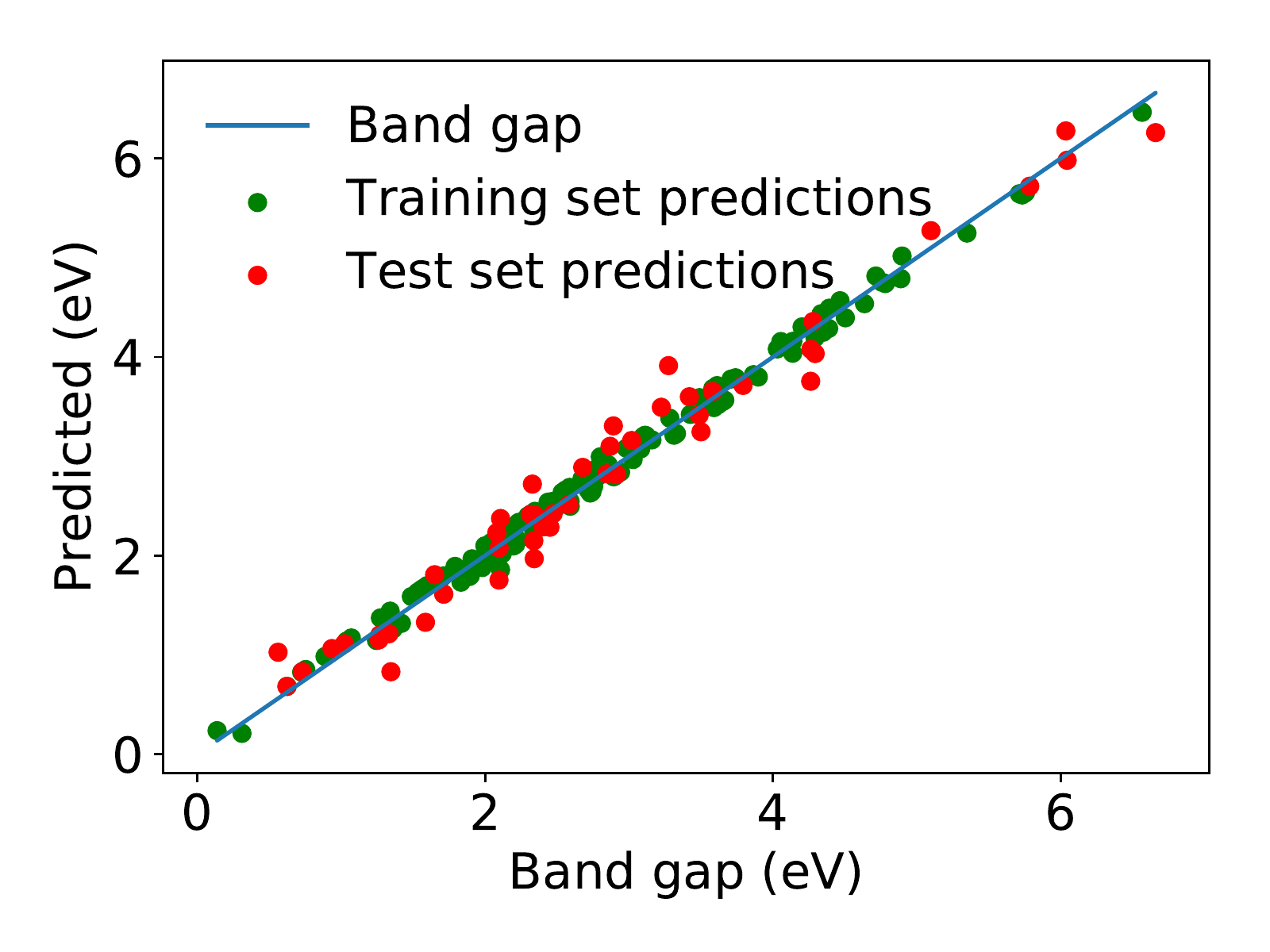}
\caption{Machine-learning predictions based on a support-vector regressor with a radial bias kernel. Predictions for the training set (green) circles and test set (red circles) are compared with the HSE calculated values (blue line).}
\label{fig2}
\end{figure}

Linear regression, which is the simplest model considered, gave a RMSE of 0.59 eV. This error is larger than desired. Support-vector regression with a linear kernel gave almost the same error. However, upon training a support-vector machine with a radial bias kernel, this error is greatly reduced; we find an RMSE of only 0.283 meV, and $R^2$ of 0.957, suggesting an excellent fit.

For the regression-tree-based methods, as expected, the simple decision tree gives the largest RMSE; the RMSE is reduced for the random forest regressor. The boosted regression tree gave the smallest RMSE of the tree-based methods. Boosting leads to a substantial improvement when compared to the simple decision tree; the boosted regressor has a RMSE of only 358 meV and $R^2$ = 0.934.\\\\

For the best model (nonlinear support vector machine), we plot the RMSE as a function of the training set size in Fig.~\ref{fig_sample_size}. To generate this plot, we performed $n$-fold cross validation, where increasing $n$ leads to an increase in the size of the training set. When the training set size is 124 (3-fold cross validation, the RMSE is 336 meV; increasing the training set size by over 30\%  to 167 (10-fold cross validation), the RMSE is 283 meV.

The error of only 283 meV for the nonlinear support-vector machine is sufficiently small to make the model predictive in nature. This error is around the uncertainty in the band gaps for high-level first-principles calculations. Hybrid functionals rely on choosing a mixing parameter that affects the calculated gap, whereas the results from nonselfconsistent $GW$ calculations are sensitive to the choice of starting wavefunctions \cite{WanPRL2006}; the error in the calculated gaps based on these approaches is typically 0.1 -- 0.3 eV. Therefore, our fitted model provides a degree of accuracy as good as the input band gaps calculated from first principles.

%The RMSE of only 80 meV is also remarkable considering the simplicity of the desciptors used in the fitting; importantly, our descriptors are purely based on the properties of the constituent elements, and we did not use band gaps calculated using low-level first-principles calculations as features. Lee \emph{et al.} predicted $GW$ band gaps and achieved a RMSE of 0.18 eV, but only when a total of 41 features were used, including the band gaps from GGA and modified Becke-Johnson DFT calculations. In many cases, particularly when dealing with a large materials-design space, it would be preferrable not to perform low-level DFT calculations in order to predict an accurate band gap; the present results show that this is indeed possible for kesterite semiconductors. %Pilania \emph{et al.} we able to achieve a RMSE of 80 meV when machine learning double perovskites upon cation mutation of the AA'BB'O$_6$ perovskites, consistent with the idea that machine learing algorithms perform better on sys

To visualize the accuracy of our band gap predictions, we plot each predicted gap as a function of the calculated HSE gap in Fig.~\ref{fig2}, using the optimized support-vector machine with a radial bias kernel. We partitioned the HSE-calculated  band gaps and features for the 184 compounds into a training and test set; approximately 75\% of the data points are used to train the optimized machine learning model, and 25\% are kept for testing. Figure~\ref{fig2} shows that the model provides highly accurate predictions for both the training and test set, when compared to the HSE calculated values.

\begin{figure}[t]
\centering
\includegraphics[trim=0.7cm 0cm 1.5cm 0cm, clip=true, width=8.6cm]{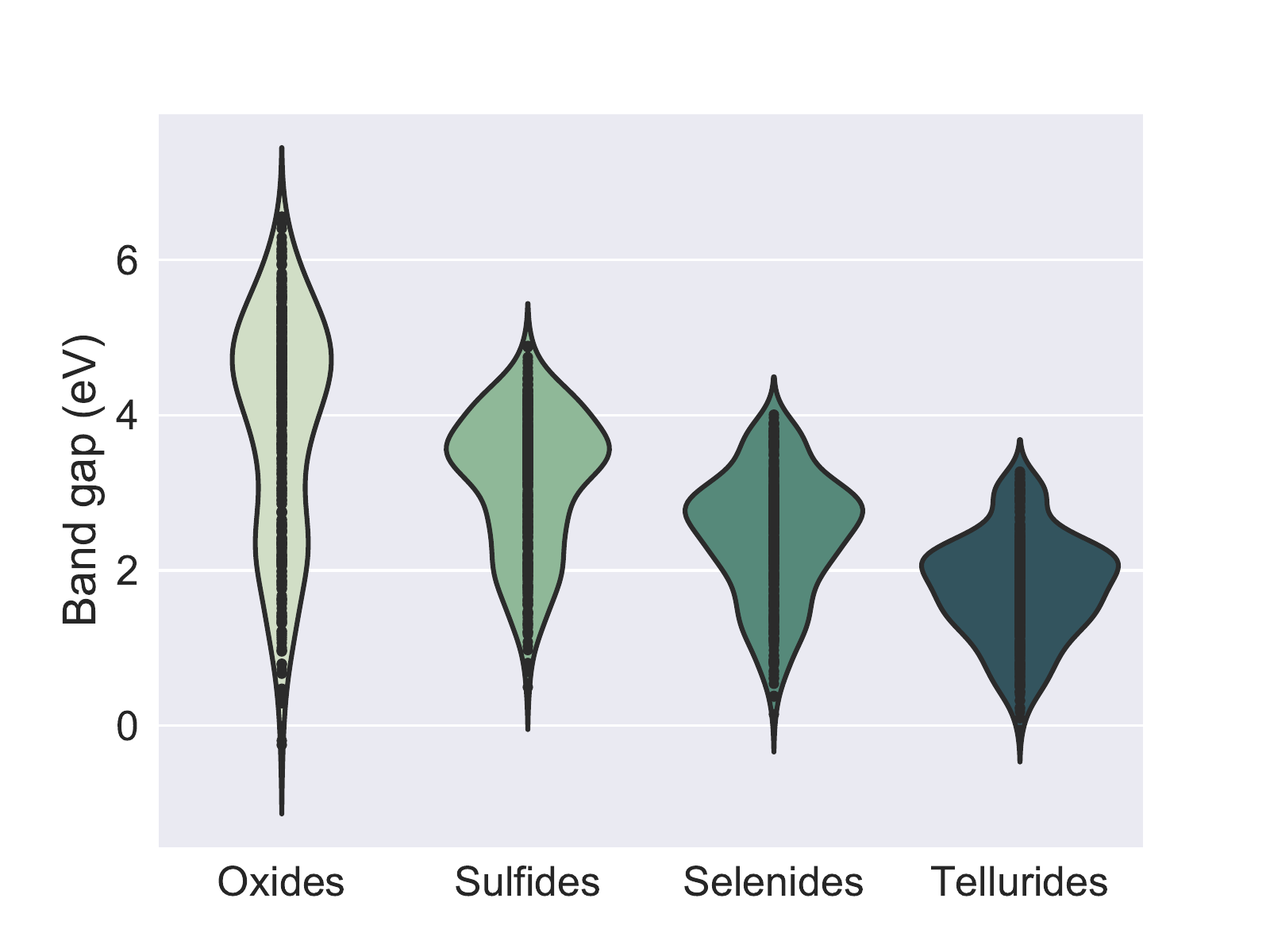}
\caption{A violin plot for the predicted band gap distributions for the oxides, sulfides,  selenides and tellurides. The width of each distribution at a given energy indicates the number of materials with a band gap around that energy.}
\label{fig3}
\end{figure}

\subsection{Direct--indirect classifier}
\label{sec:class}

Based on our HSE calculations, of the 184 gaps used for fitting, 78 were found to be direct band gaps, and 108 were indirect. The classifier is trained on the simple 12-dimensional feature space that was used in regression. The model achieves an accuracy score of 73\%. 

To provide better predictions, we preform feature engineering. We attempted to construct differences, means and standard deviations from the features, as was implemented previously \cite{LeePRB2016}; however, this did not improve classifier performance.
Improved predictions were achieved by constructing polynomial combinations of the original features in the 12-dimensional feature space. For $2nd$ order polynomial combinations, the accuracy of the classifier is increased to 83\%. Using $3rd$ order polynomial features leads to a reduction in the accuracy to 81\%; increasing the degree of the polynomial further lead to a more dramatic reduction in classifier accuracy, suggesting over fitting.

To address the issue of over fitting, while still having the advantage of keeping some higher-order terms, we use the feature-selection method with recursive feature extraction. In this way, the high-dimensional polynomial feature space is pruned to a small subset of features that have the highest weighting in determining the outcome. Using feature extraction by fitting the classifier with $3rd$ order polynomial features, the accuracy score is increased to 89\%; the optimum number of features is 30. Our optimized binary classifier is described by the following metrics for classification performance: $precision$ = 0.88,   $recall$ = 0.91, $f1$ = 0.89.

%To further increase the accuracy of the classifier, we fitted the classifier using 5th order polynomial combinations of the feature space and using recursive feature extraction. By extracting the top XX features, the classifier accuracy score was.

\begin{table}
\setlength{\tabcolsep}{14pt}
\setlength{\extrarowheight}{4.5pt}
\begin{tabular}{lccccc} \hline\hline
 Feature space   &  Accuracy score   \\ \hline
15--D & 73\%       \\
2nd--PF   &  83\%  \\
3rd--PF   &   81\%  \\
3rd--PF$+$FS  &  89\% \\  \hline\hline
\end{tabular}
\caption{Accuracy score for the direct--indirect classifier using logistic regression. Results are shown for different models with varied complexity in the feature space: (i) the simple 15-dimensional (15--D) feature space described in Sec.~\ref{sec:features}; (ii)  2nd order polynomial combinations (2nd--PF) of the 15-D set; (iii) 3rd order polynomial features (3rd--PF); and, (iv) 3rd order polynomial features plus feature extraction (3rd--PF$+$FS ).}
\label{tab:error}
\end{table}	

\subsection{Predicted results}
%\begin{table*}
%\setlength{\tabcolsep}{4pt}
%\setlength{\extrarowheight}{4.5pt}

%\label{tab:error}
%\end{table*}

\begin{figure}
\centering
\includegraphics[trim=0.0cm 0cm 1.5cm 0cm, clip=true, width=8.6cm]{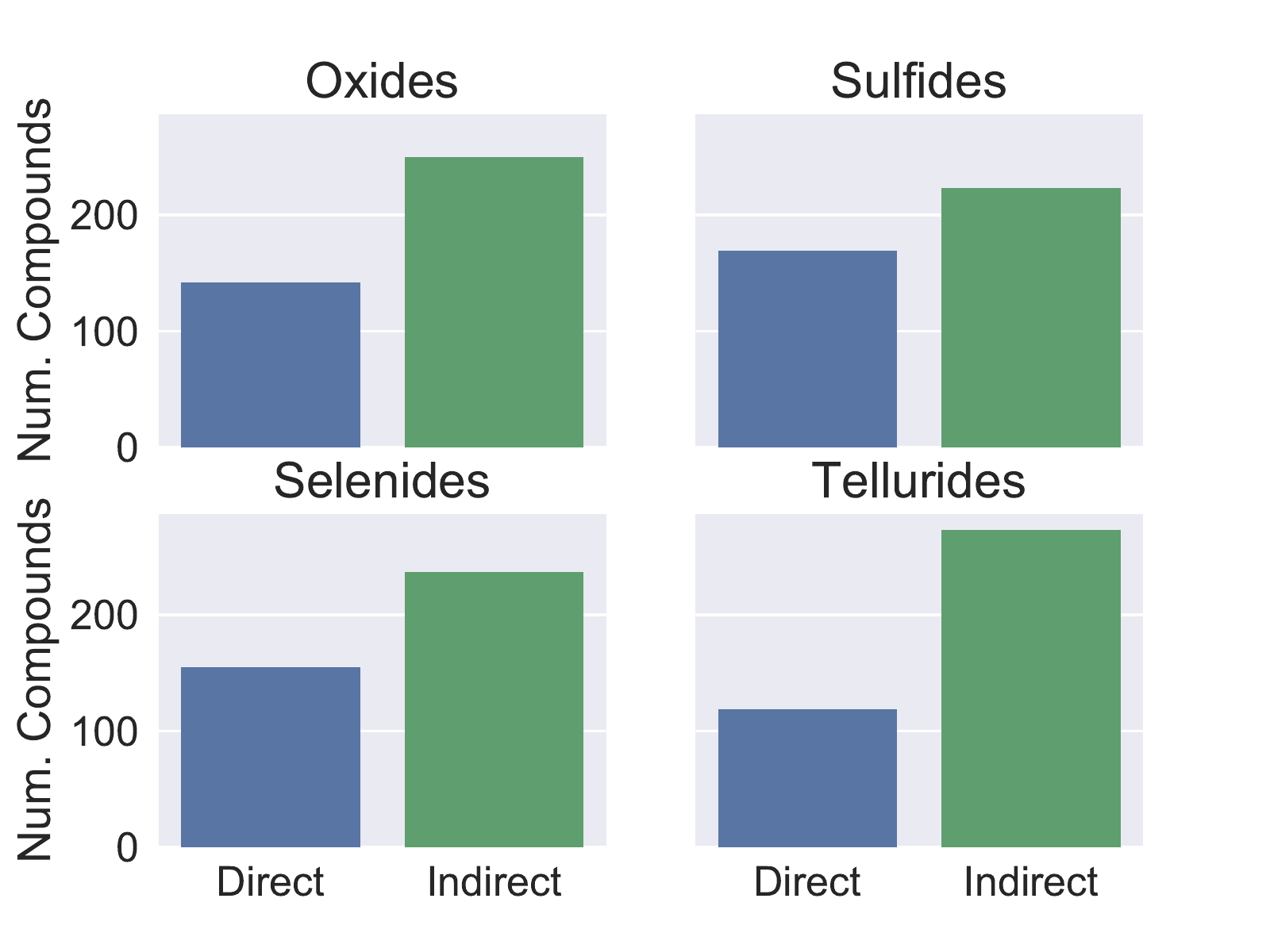}
\caption{Direct--indirect distributions for the band gaps of the oxides, sulfides,  selenides and tellurides.}
\label{fig4}
\end{figure}

\subsubsection{Band gaps}

With our fitted models, the band gaps of all 1568 materials are predicted. In Fig.~\ref{fig3}, the band gap distributions are presented for the oxides, sulfides, selenides and tellurides. Oxides typically have larger band gaps with a mean band gap $E_g^{av}$ = 3.82 eV; however, the distribution of the band gaps is over a very wide energy range, with a standard deviation of $\sigma =$ 1.47 eV. The trend moving down the periodic table for the anions is for smaller band gaps and a more localized distribution. For the tellurides, $E_g^{av}$ = 1.78 eV, and $\sigma =$ 0.68 eV.

The direct--indirect predictions are shown in Fig.~\ref{fig4}. Over all materials studied, 70\% are found to have indirect band gaps, and 30\% are direct-gap materials. The percentage of materials that were direct or indirect was anion dependent, however there was no clear systematic trend.

Materials with a band gap in the range 0.5--2.5 eV are suitable solar absorbers \cite{RuhleSE2016}; of the 1568 materials studied, 717 had a band gap in this range. For optimum photovoltaic device performance, band gaps around 1.5 eV are optimum \cite{RuhleSE2016}. We have identified 242 materials with a band gap in the ``\emph{optimum range}" of 1.2 -- 1.8 eV. The band gap properties of all 1568 kesterite I$_2$-II-IV-VI$_4$ compounds are tabulated in the supplementary material.

\subsubsection{Material stability}
%We did not make any attempt to assess the thermodynamic stability of the compounds studied in this work, and that is outside the scope of the present manuscript. It is known that kesterite materials with a large mismatch in the ionic radii of the cations tend to be unstable \cite{HongPCCP2016}. The stability of the kesterite compounds at thermodynamic equilibrium can be assessed from first-principles by constructing phase diagrams \cite{ChenPRB2010}. The stability of these materials can also be assessed in experiment by direct synthesis; moreover, nonequilibrium phases of compounds can be accessed experimentally using a number of techniques, such as reactive sputtering \cite{SunCM2017}. We encourage other researchers, both theoretical and experimental, to use the results in this manuscript as a guide in the materials engineering of kesterite I$_2$-II-IV-VI$_4$ semiconductors.

In order to guide further experimental and theoretical work, we have also assess the stability of the 242 compounds predicted to have band gaps in the  ``\emph{optimum range}" of 1.2 -- 1.8 eV. The stability was assessed by computing the enthalpy of formation for each compound, and calculating the energy of decomposition into other phases. This was achieved by making use of the Materials Project database \cite{JainAPL2013}, which contains the enthalpies of formation for hundreds of thousands of materials. In this way, we can determine whether a material is stable, metastable, or not stable. 

Of the 242 compounds, 25 were found to be the ground state for that stoichiometry with respect to the Materials Project database; i.e., these materials are expected to be stable. An additional 9 materials had an Energy Above Hull of $<$ 0.1 eV/atom, and are expected to be metastable \cite{SunSA2016}. We therefore predict that 34 of these kesterites with a band gap in the ``\emph{optimum range}"  should be synthesizable. 

\subsubsection{Band gap verification}

As a final step, we verify the machine-learned band gaps of these 34 stable compounds using first-principles calculations with the HSE functional. In the end, we find that 25 of these materials actually had a band gap with ``\emph{optimum range}" of 1.2 -- 1.8 eV.
In Table~\ref{tab:gaps},  the band gap properties for these 25 materials are presented. We indicate the magnitude of the fundamental gap, as well as the direct -- indirect character. Materials that are the ground state for that stoichiometry are indicated to have an Energy Above Hull of 0 eV. For materials that are metastable, the Energy Above Hull is indicated.

\begin{table}[H]
\setlength{\tabcolsep}{3pt}
\setlength{\extrarowheight}{4.5pt}
\begin{tabular}{lcccc} \hline\hline
Material   &  $E_g$  & Dir./Indir. & E Above Hull  &  Stability  \\
                &    (eV) &.         &   (eV) &   \\
\hline
Li$_2$BeGeTe$_4$ & 1.419 & direct & 0 & stable \\
Li$_2$BeSnTe$_4$ & 1.611 & direct & 0 & stable \\
Rb$_2$BeSnTe$_4$ & 1.692 & direct & 0 & stable \\
Rb$_2$HgTiSe$_4$ & 1.751 & indirect & 0 & stable \\
Cs$_2$HgTiSe$_4$ & 1.753 & indirect & 0 & stable \\
Cu$_2$BeSiTe$_4$ & 1.251 & indirect & 0 & stable \\
Cu$_2$BeGeSe$_4$ & 1.210 & indirect & 0 & stable \\
Cu$_2$MgSiTe$_4$ & 1.272 & indirect & 0 & stable \\
Cu$_2$SrSiSe$_4$ & 1.793 & indirect & 0 & stable \\
Cu$_2$ZnSiSe$_4$ & 1.751 & direct & 0 & stable \\
Cu$_2$ZnSnS$_4$ & 1.238 & direct & 0 & stable \\
Cu$_2$CdSiSe$_4$ & 1.534 & direct & 0 & stable \\
Ag$_2$BeSiTe$_4$ & 1.527 & indirect & 0 & stable \\
Ag$_2$BeGeSe$_4$ & 1.489 & direct & 0 & stable \\
Ag$_2$MgSiTe$_4$ & 1.591 & direct & 0 & stable \\
Ag$_2$MgGeSe$_4$ & 1.322 & direct & 0 & stable \\
Ag$_2$SrSiTe$_4$ & 1.543 & indirect & 0 & stable \\
Ag$_2$ZnSiSe$_4$ & 1.787 & direct & 0 & stable \\
Ag$_2$CdSiSe$_4$ & 1.640 & indirect & 0 & stable \\
Ag$_2$HgSiSe$_4$ & 1.218 & direct & 0 & stable \\
Cs$_2$BeSnTe$_4$ & 1.708 & direct & 0.004 & metastable \\
Na$_2$BeSnTe$_4$ & 1.783 & direct & 0.014 & metastable \\
Ag$_2$SrSnSe$_4$ & 1.272 & direct & 0.018 & metastable \\
Ag$_2$CaSiTe$_4$ & 1.720 & direct & 0.061 & metastable \\
Cu$_2$BeSnS$_4$ & 1.657 & direct & 0.086 & metastable \\
\hline\hline
\end{tabular}
\caption{Predicted properties for materials with band gaps in the  ``\emph{optimum range}" of 1.2--1.8 eV. The magnitude of the fundamental gap ($E_g$), and the direct--indirect (Dir./Indir.) character of the gap are presented. The stability is also indicated; for materials that are metastable, the Energy Above Hull (per atom) is indicated.}\label{tab:gaps}
\end{table}

The 25 materials presented in Table~\ref{tab:gaps} are largely unexplored. We encourage other researchers, both theoretical and experimental, to use the results in this manuscript as a guide in the materials engineering of kesterite I$_2$-II-IV-VI$_4$ semiconductors.

\section{Conclusions}
\label{sec:conclusions}
We have determined the band gap properties of 1568 kesterite I$_2$-II-IV-V$_4$ semiconductors using a combination of first-principles calculations and machine learning. By performing explicit hybrid-functional calculations on a subset of 200 compounds, we trained machine learning-models to predict the magnitude and character of the fundamental gap. A trained machine learning regressor based on a support-vector machine could predict the magnitude of the gap with a RMSE of 283 meV; a direct--indirect classifier was fit using logistic regression, and has an accuracy of 89\%. Our predictions identify 242 materials with a band gap in the ``\emph{optimum} range" of 1.2 -- 1.8 eV, and we expect that 34 of these materials are synthesizable; 25 of these materials actually had a band gap in the range of of 1.2 -- 1.8 eV, as verified using first-principles calculations with the HSE functional. These results will be useful in the materials engineering of solar absorbers for photovoltaic devices.

\begin{acknowledgments}
We gratefully acknowledge computational resources provided by the Australian National Computational Infrastructure (NCI), and support from the Australian Research Council.
\end{acknowledgments}

%\bibliographystyle{apsrev}
%\bibliography{ML}

\end{document}